\documentclass[compsoc,conference,a4paper,10pt,times]{IEEEtran}
\IEEEoverridecommandlockouts
\usepackage{amsmath,amssymb,amsfonts}
\usepackage{algorithmic}
\usepackage{bmpsize}
\usepackage{cite}
\usepackage{enumitem}
\usepackage{graphicx}
\usepackage[colorlinks=true,urlcolor=black]{hyperref}
\usepackage[noabbrev,capitalise]{cleveref}
\usepackage{lipsum}
\usepackage{textcomp}
\usepackage{xcolor}
\def\BibTeX{{\rm B\kern-.05em{\sc i\kern-.025em b}\kern-.08em
    T\kern-.1667em\lower.7ex\hbox{E}\kern-.125emX}}

\usepackage{multirow}
\usepackage{booktabs}
\usepackage{tikz}

\begin{document}

\title{On the Effect of Ruleset Tuning and Data Imbalance on Explainable Network Security Alert Classifications: a Case-Study on DeepCASE}

\makeatletter

\newcommand\copyrighttext{%
  \footnotesize \textcopyright 2025 IEEE. Personal use of this material is permitted. Permission from IEEE must be obtained for all other uses, in any current or future media, including reprinting/republishing this material for advertising or promotional purposes, creating new collective works, for resale or redistribution to servers or lists, or reuse of any copyrighted component of this work in other works.
  DOI: \href{https://doi.org/10.1109/EuroSPW67616.2025.00009}{10.1109/EuroSPW67616.2025.00009}
}
\newcommand\copyrightnotice{%
\begin{tikzpicture}[remember picture,overlay]
\node[anchor=south,yshift=10pt] at (current page.south) {\fbox{\parbox{\dimexpr\textwidth-\fboxsep-\fboxrule\relax}{\copyrighttext}}};
\end{tikzpicture}%
}
\makeatletter
\def\ps@headings{
\def\@oddhead{\hbox{}\@IEEEheaderstyle\rightmark\hfil\thepage}\relax
\def\@evenhead{\@IEEEheaderstyle\thepage\hfil\leftmark\hbox{}}\relax
\let\@oddfoot\@empty
\let\@evenfoot\@empty
}
\def\ps@IEEEtitlepagestyle{
\def\@oddhead{\hbox{}\@IEEEheaderstyle\leftmark\hfil\thepage}\relax
\def\@evenhead{\@IEEEheaderstyle\thepage\hfil\leftmark\hbox{}}\relax
\def\@oddfoot{\copyrightnotice}
\def\@evenfoot{\copyrightnotice}
}

\makeatother

\pagestyle{headings}
\thispagestyle{IEEEtitlepagestyle}

\author{\IEEEauthorblockN{Koen T. W. Teuwen,
Sam Baggen, Emmanuele Zambon and
Luca Allodi}
\IEEEauthorblockA{\textit{Eindhoven University of Technology}\\
Eindhoven, The Netherlands\\
\{k.t.w.teuwen, s.a.m.baggen, e.zambon, l.allodi\}@tue.nl}}

\maketitle

\begin{abstract}
Automation in Security Operations Centers (SOCs) plays a prominent role in alert classification and incident escalation. However, automated methods must be robust in the presence of imbalanced input data, which can negatively affect performance. Additionally, automated methods should make explainable decisions. In this work, we evaluate the effect of label imbalance on the classification of network intrusion alerts. As our use-case we employ DeepCASE, the state-of-the-art method for automated alert classification. We show that label imbalance impacts both classification performance and correctness of the classification explanations offered by DeepCASE. We conclude tuning the detection rules used in SOCs can significantly reduce imbalance and may benefit the performance and explainability offered by alert post-processing methods such as DeepCASE. Therefore, our findings suggest that traditional methods to improve the quality of input data can benefit automation.
\end{abstract}


\section{Introduction} \label{sec:introduction}
\textit{Security Operations Centers (SOCs)} are important to an organization's security strategy. A SOC monitors network and/or host activity to detect malicious events~\cite{knerler2022}. SOCs employ a variety of tools to aid in monitoring efforts, including \textit{Network Intrusion Detection Systems (NIDS)}, which have been known to generate an unmanageable volume of security events for suspicious activities~\cite{soc-systematic-study}.
Suricata~\cite{oisf2024} is a common signature-based NIDS that employs \textit{rules} defining the malicious behavior to detect and raise \textit{alerts}. SOC analysts use related events and contextual information to determine whether an incident has occurred~\cite{kersten-23}. A substantial part of the resulting workload can be considered unnecessary due to the high prevalence of False Positives (FPs, i.e., false alerts)~\cite{yang2024}.

Previous research improves detectors using traditional methods, such as studying effective rule design~\cite{ruling-the-unruly}. Tuning of IDS rulesets~\cite{knerler2022} is commonly applied in SOCs but can negatively affect IDS coverage~\cite{vermeer2023-alert-alchemy}.
A second stream of work reduces the workload required to process incoming alerts~\cite{vanede2022, hassan2020, wang-24}.
Similarly to the analyst process~\cite{kersten-23}, these methods leverage contextual events to determine which alerts are related to incidents.

The trustworthiness of automated solutions in SOCs depends on their robustness in light of data characteristics such as concept drift~\cite{andresini-21, barbero-22} or the emergence of previously unseen classes~\cite{han-23}, and on the explanations offered by these methods~\cite{wei-23} to help analysts understand potential cybersecurity incidents.
Although it is well known that network intrusion data is highly imbalanced~\cite{yang2024, vermeer2022-ruling-the-rules, ruling-the-unruly}, and related studies suggest that imbalanced data~\cite{ho2017, kaur2019} negatively affect Machine-Learning (ML) performance, the effect of high imbalance on the performance of alert post-processing solutions remains unclear. For example, if a malicious connection overlaps with multiple benign connections in the feature space, a classifier might classify an attack as irrelevant, favoring performance on the majority class~\cite{lopez-13}. Previous work on imbalance in the security domain has not focused on the effects of imbalance on performance itself, but rather on perceived performance through the choice of appropriate evaluations methods (i.e., how imbalance may result in misleading performance evaluations that are not representative of performance in real-world usage scenarios)~\cite{axelsson-00,pendlebury19-tesseract,andresini-21,arp2022}. Although it is reasonable to assume that imbalance impacts ML performance beyond misleading evaluations, it remains unclear what the magnitude of the impact is and how imbalance impacts performance compared to other dataset characteristics such as dimensionality~\cite{curse-of-dimensionality} or heterogeneity~\cite{liu22}. The robustness of post-processing methods on highly imbalanced SOC data remains an open problem.

In this work, we study the effect of NIDS tuning on \mbox{(semi-)automated} alert post-processing methods. One recent method to automatically filter alerts is DeepCASE, which has proven promising in initial evaluations~\cite{vanede2022}. In this paper, we take DeepCASE as a case-study to evaluate the impact of data imbalance on post-processing performance. 
Further, we also investigate the explainability of post-processing methods, the lack of which can be concerning~\cite{nadeem2023}.\\
\noindent Concretely, we make the following contributions.
\begin{itemize}[leftmargin=*]
    \item To the best of our knowledge, we are the first to study the effects of imbalance on the post-processing of security alerts using real SOC data.
    \item We find that current post-processing methods like DeepCASE might not be robust against high imbalance, worsening classification performance and explanations.
    \item Further, our findings suggest that tuning is effective in improving the quality of data input to these methods, increasing classification performance and explainability.
\end{itemize}

First, we discuss the relevant background and related work in \cref{sec:background_related_work}. Then, we present the methodology of our study in \cref{sec:methodology}, and our findings in \cref{sec:results}. Finally, we discuss the implications and limitations of our findings in \cref{sec:discussion} and conclude our work in \cref{sec:conclusion}.

\section{Background and Related Work} \label{sec:background_related_work}
In \cref{sec:deepcase}, we discuss the background on DeepCASE, the topic of our case-study on alert post-processing in SOCs, and discuss related work on imbalance and machine learning for intrusion detection in \cref{sec:related}.

\subsection{Background on DeepCASE} \label{sec:deepcase}
DeepCASE~\cite{vanede2022} belongs to the group of methods to reduce SOC workloads through post-processing using ML methods, and is designed to classify alerts to distinguish attacks from FPs by evaluating alerts in the \textit{context} of preceding alerts regarding the same host (victim). To this end, it uses two components: a context builder to encode alerts and their context together into a \textit{sequence}, and an interpreter to cluster similar sequences. Clusters are manually labeled, and new alerts close to an existing cluster are automatically assigned the cluster label. The behavior of both components can be tuned using hyperparameters described in \cref{tab:2hyper} of Appendix~\ref{app:hyper}.
DeepCASE was evaluated on the Lastline dataset, consisting of $10.5M$ NIDS alerts from a real-world SOC, with $45.1k$ attacks.

The context builder is based on a Recurrent Neural Network (RNN) with attention mechanism~\cite{attention-mechanism}, and utilizes a sliding window to collect the context of each alert. Preceding alerts are selected based on the involved (victim) host and only within a fixed time window. The events in the resulting sequence are represented by one-hot encoding. The context builder then derives an \textit{attention vector} from the sequence that estimates how relevant the context events are. To derive this attention vector, the RNN is trained to predict alert $e_i$ given its context $[e_0, \hdots, e_{i-1}]$. A confidence score associated with this prediction is used as a parameter to decide whether the attention vector is reliable enough to proceed with the classification.

After creating the attention vectors for the sequences, DeepCASE's interpreter clusters the sequences together using both the attention vector and the context, which are combined into a \textit{vector of total attention}. The interpreter uses DBSCAN~\cite{ester1996} to cluster these vectors.
After clustering, a manual step is taken in which the clusters are labeled by evaluating a sample of sequences assigned to the cluster. Thereafter, a cluster and all sequences assigned to it automatically receive the label of the sequence with the highest risk.
We note that similar contexts result in similar vectors of total attention per event, which implies that DeepCASE implicitly assumes that the contexts have similar relevant contextual events and thus are likely to have the same output label.

In addition to the labels assigned during the manual cluster analysis, alerts can be classified into one of three rejection labels. Firstly, it is possible that the confidence of the context builder is insufficient to use a resulting attention vector for clustering. Secondly, a sequence may contain events of a type that was not present in the training data. Lastly, vectors of total attention may be too far from any cluster to be assigned a label. In all of these cases, an analyst must manually classify the alert.

The attention vectors derived from the context builder are argued to be useful for explaining DeepCASE's automatic classification of alerts~\cite{vanede2022}, providing insight into DeepCASE's internals. Nadeem \textit{et al.} present a SoK study on explainable AI for cybersecurity~\cite{nadeem2023}. One of their takeaways is that involving humans in the evaluation of the quality of explanations is crucial, and that only $14\%$ of the reviewed literature performs such user studies. Before our work, no evaluation of the correctness of the explanations offered by DeepCASE had been carried out.

\subsection{Related work} \label{sec:related}
Several definitions for label imbalance and related metrics exist, such as entropy. For this study, we use the definition of \textit{Imbalance Ratio (IR)} from Lorena \textit{et al.}~\cite{lorena2019} that generalizes to multi-class settings and is limited to the interval $[0..1)$ The definition of IR is given in \cref{eq:IR} where $n_c$ is the number of classes, $n_{c_i}$ is the number of samples in class $i$, and $n$ is the number of samples in the complete dataset. An IR close to $1$ indicates an imbalanced dataset, and an IR close to $0$ indicates a balanced dataset. To characterize typical IDS data, we note that the dataset described by Yang \textit{et al.}~\cite{yang2024} has an IR of $1-4\mathrm{e}{-6}$.

\vspace{-0.05in}
\begin{equation}
    IR = 1 - \frac{1}{\frac{n_c - 1}{n_c} \sum_{i = 1}^{n_c} \frac{n_{c_i}}{n - n_{c_i}}}
    \label{eq:IR}
    \vspace{-0.05in}
\end{equation}

Recent work focuses on NIDS explainability~\cite{wei-23} and robustness, specifically against concept drift~\cite{andresini-21, barbero-22} (i.e., changes in data distributions) that can affect performance. Some of these changes, such as those caused by adversarial behavior, can be considered inherent to the context in which IDS operate, like imbalance is to the context in which post-processing methods operate~\cite{yang2024}. \textit{Label imbalance} is a phenomenon in which the relative size of data classes differs significantly~\cite{lorena2019} and can negatively affect ML performance~\cite{ho2017, kaur2019}. Similarly, other data(set) characteristics such as dataset size~\cite{prusa-15}, heterogeneity~\cite{liu22}, and dimensionality~\cite{curse-of-dimensionality} can affect performance. In this work, we study the robustness of post-processing methods against label imbalance and also relate this to the quality of explanations.

From the literature, we identify two semi-supervised methods with a purpose similar to DeepCASE: \textit{(1)} NoDoze~\cite{hassan2019}, and \textit{(2)} AlertPro~\cite{wang-24}.
NoDoze ranks alerts and relies on the assumption that rare events are anomalous. Using contextual information, NoDoze constructs large dependency graphs that include many FPs and thus must merge paths with similar anomaly scores to reduce the number of irrelevant paths retrieved.
AlertPro uses a variety of features, some of which are extracted from the context (i.e., surrounding alerts) to estimate the risk associated with alerts. 
Previous research on SOC workflows~\cite{vermeer2023-alert-alchemy} has proposed that post-processing research could benefit from improving or tuning detectors, as this would improve the quality of the data used to train alert classifiers. Since the tuning of rulesets involves disabling noisy rules whilst not affecting low-noise rules, the proportion of generated noise, and therefore the imbalance, is reduced. In this study, we explore the effects of tuning as a method for reducing imbalance and benefiting post-processing methods.

Past research has examined work on ML methods and datasets in the security domain and found several bad practices that threaten the validity of research results.
Arp \textit{et al.}~\cite{arp2022} identified common pitfalls of ML in computer security, of which most can affect the validity of evaluations. Two notable pitfalls are spurious correlations and data snooping. \textit{Spurious correlations} refer to a phenomenon by which features unrelated to the security problem (e.g., resulting from the mixing of malicious and benign data from different origins) artificially inflate the performance of ML to be unreflective of real-world performance. \textit{Data snooping} is a pitfall that refers to ML training using data that would normally be unavailable for training (i.e., incorporation of test data into the training set) and can lead to inflated performance due to unrealistic `foreknowledge'. They also mention the base rate fallacy, stressing the importance of using metrics robust against label imbalance for performance evaluation and their interpretation but do not focus on the fundamental effects on performance if suitable metrics were appropriately interpreted. Other previous research also discussed the impact of imbalance on perceived performance~\cite{axelsson-00,pendlebury19-tesseract,andresini-21}, but, to the best of our knowledge, the impact on the performance itself has not been studied yet, especially in the context of ML approaches for alert post-processing in SOCs. Flood \textit{et al.}~\cite{flood-24} focus specifically on benchmark dataset designs and their application. Liu \textit{et al.}~\cite{liu22} conducted a similar study on error prevalence in NIDS datasets and found several problems in commonly used datasets. Among the issues highlighted by all these works, a common concern is the suitability of the data (i.e., realism, variety, volume, correctness). In this work, we account for relevant pitfalls from these works and explore effects of label imbalance on the performance itself using suitable metrics.

\section{Methodology} \label{sec:methodology}
In order to study the effects of imbalance on the classification performance and explainability of alert post-processing methods, we devise an experimental methodology in which we build datasets that vary in imbalance using data from a cooperating SOC and compare the performance of DeepCASE on those datasets.\\
\noindent We address the following research questions.
\begin{itemize}[leftmargin=2.5em]
    \item[\textbf{RQ1}] \emph{How does label imbalance affect the classification performance of DeepCASE in the context of the cooperating SOC?}
    \item[\textbf{RQ2}] \emph{How does label imbalance affect the correctness of the explanations offered by DeepCASE?}
\end{itemize}
Our methodology accounts for several of the bad practices described in \cref{sec:related} to mitigate adverse effects on the validity of the research.

\subsection{Data Provisioning} \label{sec:data_provisioning}
The cooperating SOC, monitoring several organizations using a combination of network- and host-based intrusion detection systems, has made their data available for this study\footnote{Ethical considerations regarding the data handling have been pre-approved by the research institution at the time of the instantiation of the collaboration with the involved SOC and include usage of \mbox{(pseudo-)anonymized} data to prevent identification of individuals.}.
The data shared were collected from the network of an educational institution. The monitored assets include institutional desktop computers and servers, as well as personal devices in a Bring-Your-Own-Device (BYOD) setting. We focus on Suricata~\cite{oisf2024} alerts the SOC collected from network traffic.
\footnote{A comparison with the Lastline dataset employed in DeepCASE is provided in Table~\ref{tab:datasets} of Appendix~\ref{app:reproduce}.}

The SOC dataset consists of $24.5M$ alerts generated by $489$ unique Suricata rules. It was collected over a period of approximately five weeks between March 29 and May 4, 2022.
The alerts collected involve approximately $1.3k$ distinct hosts in the monitored environment.
Due to the lack of incidents during the collection period, none of the alerts generated by the rules are related to security incidents. Therefore, we use the same dataset as \cite{kersten2023}, in which traffic from $10$ replayed successful attacks is integrated in the SOC dataset, generating $616$ alerts.
The traffic captures are replayed in the SIEM environment in such a way that they are indistinguishable from the live traffic, in accordance with existing recommendations for traffic injection~\cite{engelen2021}.
Concretely, injected traffic consists of both benign and malicious traffic that is mapped to the same IP range as the existing traffic in the environment. The introduction of this traffic is analogous to the introduction of a new device in the BYOD setting from which the SOC dataset is collected. All protocols present in the injected traffic are also present in the SOC dataset. The injection results in the same alerts raised by the signature-based IDS as would otherwise have been raised if the attack and benign traffic were exhibited by a host in the environment where the SOC dataset was collected. We manually asserted that no alerts were raised that we believe to be the result of the manner in which the attacks were injected by inspecting every alert raised alongside the corresponding attack PCAP and verifying that its trigger conditions are explainable by the nature of the attacks and not due to the traffic not being part of the monitored network. We also verify that the alert timestamp coincides with the data collection period of the SOC dataset, and the IP addresses are consistent with the nature of the injected incident and the environment whose data should be matched. Note that DeepCASE is not concerned with low-level details of the injected traffic since it operates on the level of the alerts, which describe high-level security events in relation to hosts.
After injection, the $616$ events are manually inspected and all found to be related to the injected (successful) attack scenarios.
The collaborating SOC uses only two labels for the data: \textit{\texttt{Non-Incident}} and \textit{\texttt{Incident}}, where \texttt{Incident} alerts are alerts related to successful attacks which should be escalated and \texttt{Non-Incident} alerts are all alerts considered FPs or otherwise not worth escalating.

\subsection{Experiment Design}\label{sec:experiment_design}

\noindent\textbf{Data preparations}
To prepare DeepCASE for the SOC dataset, we perform hyperparameter optimization using the first $1\%$ of the sequences, similar to the original work~\cite{vanede2022}. \footnote{Details are provided in Appendix~\ref{app:hyper}. We reproduce DeepCASE results in ~\cite{vanede2022} with our dataset as described in Appendix~\ref{app:reproduce} and confirm that the performance of DeepCASE is similar to the performance reported for the Lastline dataset. We also perform a control experiment (see \cref{sec:control_experiment_hyperparameters}) using the default DeepCASE hyperparameters to verify that the hyperparameter optimization was successful.} The $99\%$ of sequences chronologically following the optimization set is used as the experiment set. The experiment set is further split into a train and a test set, with $20\%$ of the data in the train and $80\%$ in the test set.
To increase DeepCASE chance at learning and classifying the \texttt{Incident} alerts, we copy to the training set some sequences from the \texttt{Incident} class in the test set. DeepCASE would otherwise struggle since all \texttt{Incident} alerts would be in the test set. To this end, we randomly sample sequences from the test set with the \texttt{Incident} label and copy them to the train set until the percentages of \texttt{Incident} alerts in both sets are similar. We consider this to be permissible since the primary goal of this work is not to assess the performance of DeepCASE w.r.t. the original evaluation, but to assess the impact of imbalance on its performance, which is not obscured.
Although the incorporation of incidents from the test set into the training set constitutes a form of temporal data snooping that may artificially inflate performance on the minority class, this is equally present in all experiments we compare and therefore does not inhibit our ability to study the effects of imbalance on DeepCASE. We remind the reader that the goal of this research is not to suggest the expected level of performance of DeepCASE when deployed in real-world environments.

To study the effects of imbalance, we require datasets that vary in IR.
We create tuned datasets by simulating the tuning process commonly employed by SOCs~\cite{vermeer2023-alert-alchemy}: SOCs commonly disable noisy rules to suppress alerts generated by those rules. We reproduce various versions of the rule filter that the SOC implemented after data collection. To this end, we deactivate rules retroactively inline with the SOC filter, and drop all alerts associated with selected rules from the tuned datasets. According to the responsible SOC engineers, the rule filter became more refined over time, hence making datasets from more recent filters (High IR + Medium IR) include fewer FPs. We generate an additional dataset (Low IR) using more aggressive filtering, whereby per-host rule filters are changed to global rule filters to filter a rule not only for a single host but for all hosts. Both filter methods are normally applied by the SOC.
Critically, rule filtering is a process that automated methods such as DeepCASE should be able to cope with, since rule filtering is a common low complexity practice for SOCs to reduce analyst workloads and is considered best practice in the literature~\cite{knerler2022}.
All tuned datasets have a label IR of a different order of magnitude, varying between $1-5\mathrm{e}{-5}$ for the unfiltered dataset and $1-2\mathrm{e}{-2}$ for the low IR dataset.

\noindent\textbf{Effects on classification performance} \label{sec:experiment_design_classification_performance}
The NIDS rule adjustment process described in \cref{sec:data_provisioning} results in three datasets to evaluate the effect of the rule adjustment process on classification performance. However, they alone are insufficient to attribute any performance difference solely to label imbalance, because tuning also affects other dataset characteristics that may simplify decision boundaries. Therefore, we build four control datasets in which we vary other characteristics that may affect performance, to determine whether imbalance is indeed more influential on classification performance than these other characteristics (RQ1).
We determine which characteristics to control for by: \textit{(i)} collecting a preliminary list from the existing literature (see \cref{sec:related}) and our understanding of DeepCASE; \textit{(ii)} completing the list by interviewing authors of DeepCASE and \textit{(iii)} ensuring that the identified characteristics are indeed affected by the tuning process.
The four characteristics resulting from this process are: \textit{(1)} dataset size~\cite{prusa-15}, \textit{(2)} event imbalance (i.e., classes are defined by event type instead of label), \textit{(3)} context heterogeneity~\cite{liu22} (i.e., number of unique context vectors in the data) and \textit{(4)} context builder input dimensionality~\cite{curse-of-dimensionality} (i.e., number of unique event types).
In an effort to prevent spurious correlations~\cite{arp2022}, we construct datasets by retroactively removing rules, sequences, or alerts to vary the identified factors, rather than injecting additional events.
Concretely, we randomly sample data points to remove according to varying probabilities for each control dataset. These probabilities are derived from dataset statistics (i.e., alert/context frequencies) to influence a subset of all identified factors whilst maintaining others at levels similar to those of the unfiltered or tuned datasets. We generate control datasets for every tuned dataset to match some of its properties and vary other characteristics.
These control processes and their effects on the identified characteristics are further detailed in Appendix~\ref{app:control}.
To distinguish the effects of each controlled characteristic, we run DeepCASE on the control datasets for each of the high, medium, and low IR. Each run uses the hyperparameters obtained from the hyperparameter optimization described in Appendix~\ref{app:hyper}. Each run is executed five times to account for the stochasticity of neural network outputs.
Since the control datasets vary label imbalance, dataset size, event imbalance, context hetetogeneity, and context builder input dimensionality in manners uniquely distinct from the tuning process, these allow us to attribute performance improvements to specific properties of the dataset such as imbalance.

To analyze the effect of the IR on classification performance, we use a combination of descriptive statistics and regression analysis. We use a robust linear model to assess which characteristic most significantly affects classification performance. To account for scale differences between variables, all variables are scaled to have mean $0$ and standard deviation $1$. To ensure that no collinear variables are used in the regression, we compute the Variance Influence Factor (VIF) and remove any variables with a VIF greater than 10~\cite{slinker1990, maddala1992, gray1994} from the regression. We report statistical significance for each estimated effect size from a t-test; p-values below $0.05$ are considered significant. In addition, we use a Kolmogorov-Smirnov (KS) Test as a goodness-of-fit test to understand how much variance can be explained by the model~\cite{massey1951}. Rejection of the null hypothesis ($H_0$) of a KS-test implies poor fit. The KS test is chosen over the usual $R^2$ due to its robustness against outliers.

\noindent\textbf{Effects on correctness of explanations} \label{sec:experiment_design_explainability}
To assess the correctness of DeepCASE's explanations (RQ2), we asked an expert from the collaborating SOC to indicate relevance of each contextual event (range $[0,1]$) to the investigated alert (i.e. manually replicating DeepCASE's `explanation' of its classification).
Considering the large size of the dataset, we build this baseline only for \texttt{Incident} alerts, labelling hundreds of contexts. To validate the correctness of the expert-labeled vectors, a researcher executes the same task, after which we validate the vectors of the expert using the secondary labeling conducted by another expert, as described in Appendix~\ref{app:expert}. Both experts rely solely on the same sequences of alert messages that are presented to DeepCASE but may interpret them using tacit knowledge~\cite{cho20} from their experience working with SOCs. After confirming high agreement between the experts, the labeling of the first expert is kept as ground truth to evaluate DeepCASE.

We compare the expert explanations with DeepCASE's explanations for the unfiltered and tuned datasets. By examining the cumulative distributions of similarities between explanations of different raters (i.e., DeepCASE and an expert) over these datasets, we can infer the effect of label imbalance on the correctness of DeepCASE's explanations. We group individual runs belonging to the same dataset together to obtain a single distribution of similarities for each dataset/rater. Note that we do not leverage the control datasets to assess the effect on explainability since the sequences manually labeled by experts can include different events and require a new manual labeling for every run of a control dataset, leading to an unreasonable amount of work.

\subsection{Metrics} \label{sec:metrics}
\noindent\textbf{Classification performance metrics}
In the original work \textit{precision} and \textit{recall} (PR) are used to evaluate, respectively, how many classifications are correct and how many classified alerts are classified correctly. The F1-score combines precision and recall into a single metric using a harmonic mean. One common method to calculate PR scores is \textit{micro-averaging}, in which equal weight is assigned to every alert. Alternatively, when \textit{macro-averaging}, the PR scores are computed separately for each class, after which the per-class scores are combined with equal weight. The macro F1-score is a more informative metric in highly imbalanced environments where the minority class performance is (more) important than the majority class performance. Thus, we use the macro F1-score for hyperparameter optimization to encourage DeepCASE to perform well on the minority class.

To reproduce and compare with the original results, we follow the same approach when calculating the macro F1-score. However, we observe that rejected alerts impact SOC operations as those alerts must be analyzed manually, and hence contribute to workload. Therefore, we develop an alternative metric, the \textit{relaxed} F1-score, to capture this. The \textit{relaxed} precision and recall, respectively, evaluate how many alerts are correctly presented and how many interesting alerts are presented to an analyst. Contrary to the original work, we consider rejected \texttt{Incident} alerts as True Positive (TP) (because worthy of analysis from a human analyst) and consider rejected \texttt{Non-Incident} alerts as FP (because they waste analyst time). For all other possible alerts/classifications, the definition of TP/FP remains identical to that from the original work~\cite{vanede2022}. Note that we do not change how the PR is computed, but only the definition of what constitutes a TP/FP to better capture the alert classification's impact on the analyst workload. As a result, rejected \texttt{Non-Incident} alerts reduce precision, and rejected \texttt{Incident} increase precision and recall. The relaxed F1-score therefore more accurately reflects operational impact of post-processing and is therefore used to assess the effects of data characteristics like imbalance on the classification performance (e.g., regression).

\noindent\textbf{Correctness of explanations metrics}
To assess the correctness of explanations, we need to compare explanations (i.e,, vectors of total attention) from different raters (e.g., expert or DeepCASE). In line with common practice, we use cosine similarity to compare vectors due to the well-understood semantics of the cosine similarity, resulting in a score in the range $[0,1]$\footnote{The minimum cosine similarity is $0$ instead of $-1$ because the vectors of total attention are non-negative and thus cannot be antiparallel.}~\cite{connor2016}.
If the cosine similarity is close to $1$, then the vectors of total attention are similar, and raters agree on the correct explanation. If the cosine similarity is close to $0$, then the vectors indicate different relevant events, and raters disagree on the explanation.

\section{Results} \label{sec:results}
We discuss the effects on classification performance in \cref{sec:results_classification_performance} after which we turn to the effects on the correctness of the explanations in \cref{sec:results_explainability}.

\subsection{Classification performance} \label{sec:results_classification_performance}

\begin{figure}[tb]
    \centering
\vspace{-0.15in}
    \includegraphics[width=\linewidth]{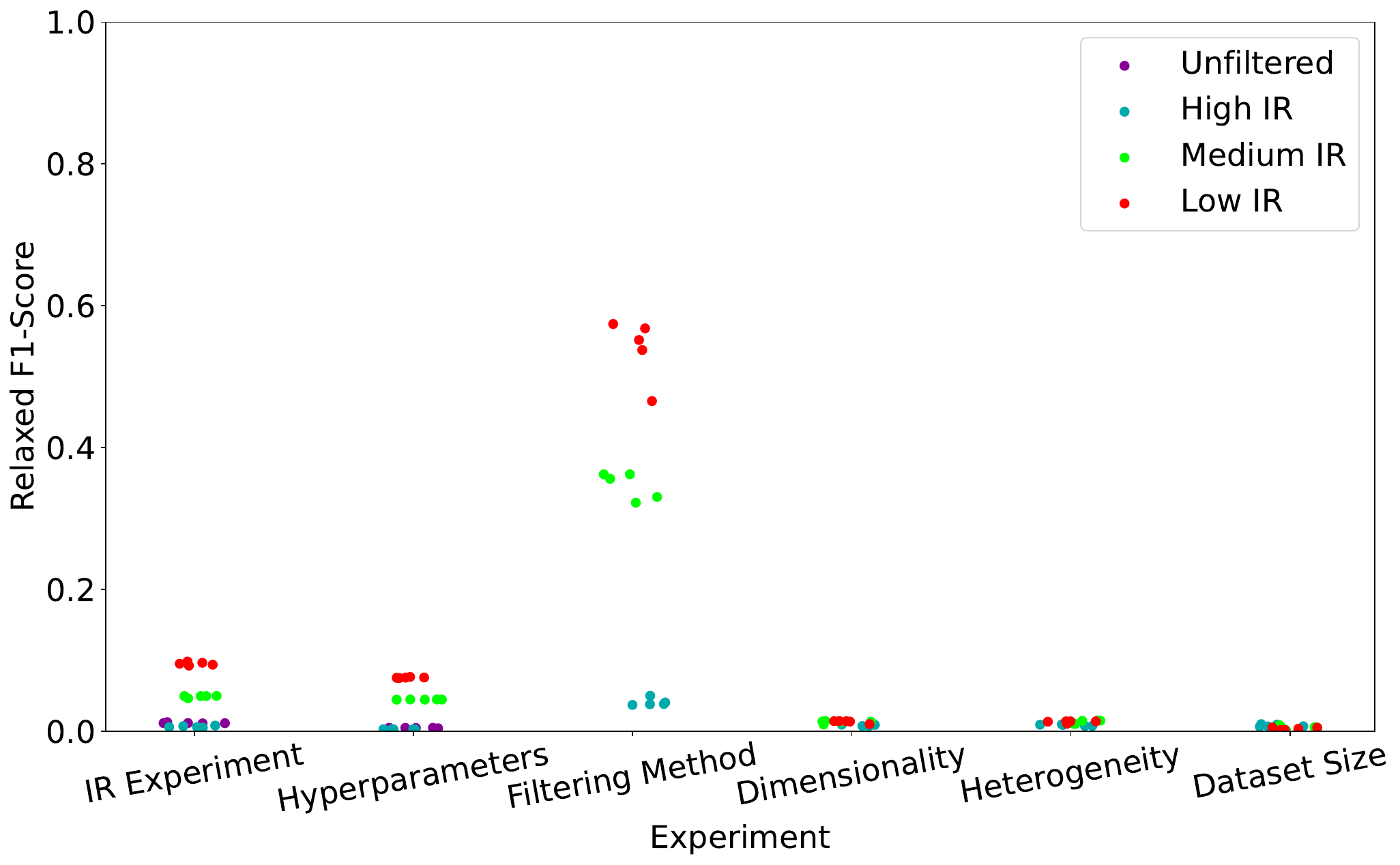}
\vspace{-0.30in}
    \caption{Relaxed F1-score for all experiments. Each point represents the relaxed F1-score for one run.}
\vspace{-0.05in}
    \label{fig:5f1}
\end{figure}

\cref{fig:5f1} shows the relaxed F1-score. Our first observation is that the performance of DeepCASE for the tuned datasets (IR Experiments) is higher than that in the hyperparameter control experiment, especially for the low IR dataset. This suggests that hyperparameter optimization was indeed successful in tweaking DeepCASE's performance on alternate data. For the various tuned datasets, and also for most control experiments, we observe a clear pattern by which DeepCASE obtains higher relaxed F1 scores on Low- and Medium-IR datasets, suggesting that a lower IR, as a result of enhanced ruleset tuning, leads to higher classification performance. \cref{tab:cir} (Appendix~\ref{app:additional}) shows the confusion matrix for the unfiltered and tuned datasets. From this confusion matrix, we can see that the Low- and Medium-IR datasets succeed in retrieving more relevant \texttt{Incident} alerts, and more importantly, retrieve less irrelevant \texttt{Non-Incident} alerts, which for the High-IR and unfiltered datasets were shown more often to analysts because they were rejected by DeepCASE.

One notable outlier is the performance on the Filtering Method control dataset. We attribute this outlier to a significant decrease in heterogeneity due to the filtering approach. Upon inspection of the confusion matrix, shown in \cref{tab:cfm} (Appendix~\ref{app:additional}), we observe that DeepCASE rejects fewer \texttt{Incident} alerts, likely due to a decrease in data complexity that makes the problem addressed by DeepCASE less complex. This decrease in data complexity is further detailed in Section~\ref{sec:filtering_method_comparison} (Appendix~\ref{app:control}). Concretely, the control dataset has a lower heterogeneity and dimensionality compared to the tuned datasets.
We note that the performance of DeepCASE on the control datasets is not necessarily representative of the performance that could be obtained by DeepCASE in practice, but it does serve a purpose in assessing the effect of the characteristics of the data on the classification performance.

To attribute differences in performance to imbalance, we use regression over the various controlled characteristics following the methodology described in \cref{sec:experiment_design_classification_performance}. The outcome is shown in \cref{tab:5reg}. The event IR, was dropped from the regression due to its VIF exceeding the threshold we set.
The test statistic of the KS-test is $0.13$ and the corresponding p-value is $0.15$. Given the p-value, we fail to reject $H_0$ and lack indications that the model poorly explains the performance differences between the various datasets. The test statistic indicates that the cumulative distribution function resulting from the regression is at most $0.13$ off from the cumulative distribution function of true relaxed F1-score, suggesting a non-negligible portion of the effect is successfully explained by the model.

\begin{table}[t]
    \centering
\vspace{-0.05in}
    \caption{Regression results showing the effect size of various characteristics on the relaxed F1-score using a Robust Linear Model.}
\vspace{-0.1in}
    \begin{tabular}{lrrr}
        \toprule
         Dimension      & Coefficient   & Confidence Interval    & $P$-value \\
         \midrule
         Constant & $0.09$ & $[\phantom{-}0.07, \phantom{-}0.10]$ & $\mathbf{<0.01}$ \\
         Label IR       & $-1.03$       & $[-1.04, -1.02]$        & $\mathbf{<0.01}$ \\
         Dataset Size   & $-0.01$       & $[-0.01, \phantom{-}0.03]$        & $0.24$ \\
         Heterogeneity  & $-0.04$       & $[-0.06, -0.03]$        & $\mathbf{<0.01}$ \\
         Unique Events  & $0.01$        & $[-0.01, \phantom{-}0.03]$        & $0.27$ \\
         \midrule
         N.obs.         &               &               & $80$ \\
         KS Test        & $0.13$        &               & $0.15$ \\
         \bottomrule
     \end{tabular}
\vspace{-0.2in}
    \label{tab:5reg}
\end{table}

The significant regression coefficients inform on characteristics that significantly influence the relaxed F1-score. We conclude the label IR appears to have the largest impact with a coefficient of $-1.03$, compared to the other characteristics we control for. Hence, the relaxed F1-score is higher for datasets with a lower label IR. The rest of the characteristics appear to have little or inconclusive effect on the relaxed F1-score, with coefficients between $-0.04$ and $0.01$. However, heterogeneity appears to have a small but significant effect, where a reduction in heterogeneity leads to a small improvement of the relaxed F1-score. To contextualize this, we note that the effect size of context heterogeneity on DeepCASE's performance is less than $5\%$ of the effect size of label imbalance.

\subsection{Explainability} \label{sec:results_explainability}
\begin{figure}[tb]
    \centering
\vspace{-0.10in}
    \includegraphics[width=\linewidth]{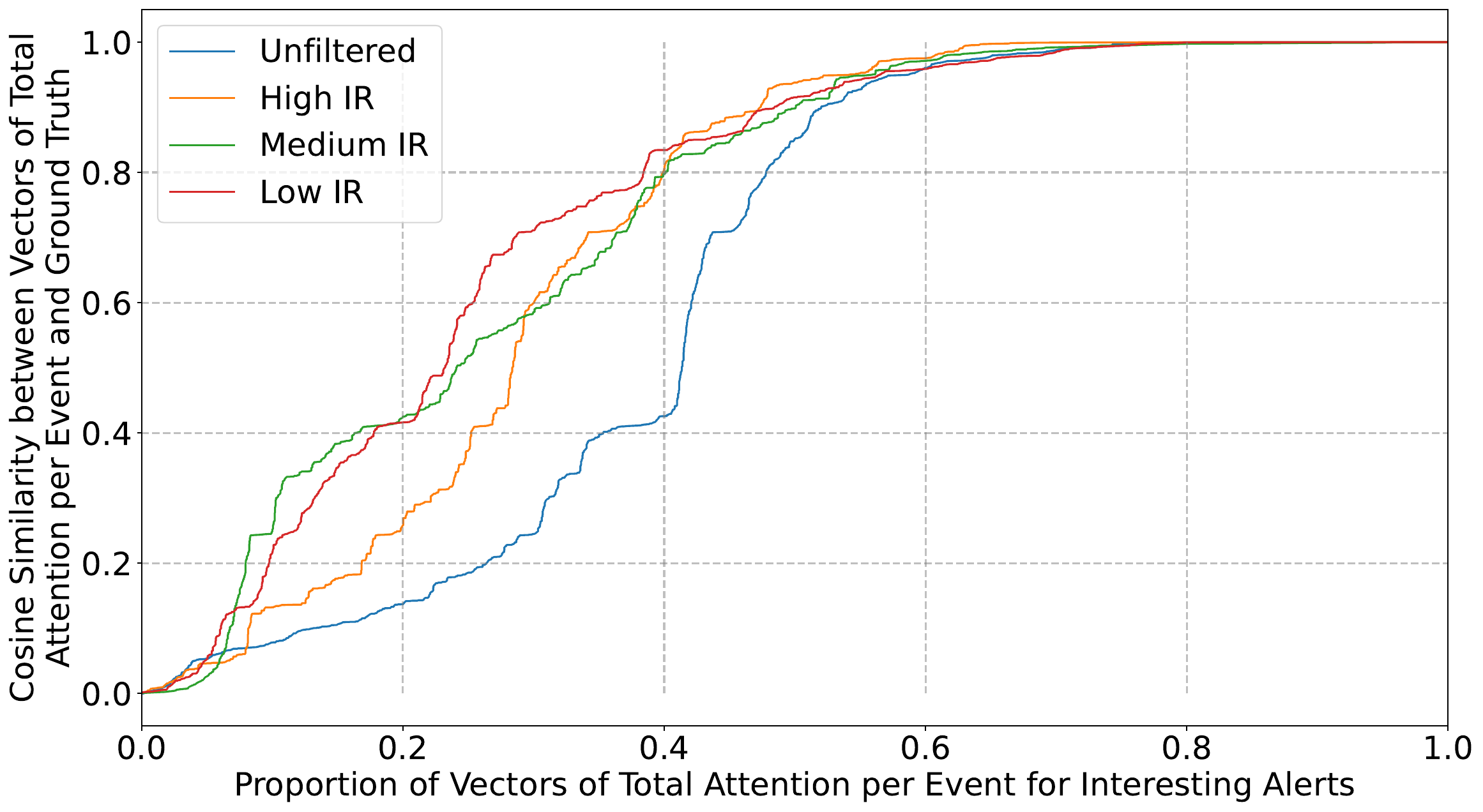}
\vspace{-0.25in}
    \caption{Cumulative distribution plots of the cosine similarity comparing of explanations from DeepCASE and the expert for the unfiltered and tuned datasets.}
\vspace{-0.15in}
    \label{fig:6cumulativeir}
\end{figure}

The cumulative distributions of the cosine similarities between the vectors of total attention per event provided by DeepCASE and the ground truth for the tuned datasets are shown in \cref{fig:6cumulativeir}.
We see that approximately $25$\% of the vectors of total attention per event is identical to the ground truth and about $50$\% of the vectors per event have a cosine similarity of $0.75$ or higher, suggesting high similarity. However, about $30$\% of the vectors have a cosine similarity of $0.25$ or lower for the unfiltered dataset, suggesting DeepCASE provides incorrect explanations for these alerts.

As we compare the unfiltered set to the tuned sets with different IR, we observe that the vectors provided by DeepCASE on the tuned datasets are usually more similar to the expert rating. Interestingly, the vectors from the Low-IR dataset also appear to outperform those from the High-IR dataset, which was also tuned to some extent. Hence, we conclude that ruleset tuning effectively assists DeepCASE in providing correct explanations.

\section{Discussion} \label{sec:discussion}
We focus on DeepCASE as a case-study for the broader class of alert post-processing methods in the context of SOCs. To support this, we compare DeepCASE with two related methods: \textit{(1)} NoDoze~\cite{hassan2019}, and \textit{(2)} AlertPro~\cite{wang-24}.
DeepCASE, NoDoze, and AlertPro all use information from surrounding alerts, similar to what analysts would do~\cite{kersten-23}.
NoDoze employs mitigation methods, such as merging paths with similar anomaly scores, to reduce the number of irrelevant paths retrieved using a threshold to decide on the similarity of anomaly scores.
Similarly to DeepCASE, where frequently occurring alerts can evict relevant alerts from contexts, in AlertPro excessive alerts can suppress relevant information from other alerts in the analyzed contexts.
Thus, both NoDoze and AlertPro rely on methods that are prone to the effects of label imbalance, similar to DeepCASE.

\noindent\textbf{Implications.}
Although it is well known that imbalance can negatively affect ML performance~\cite{santos-23}, prior to this work it was unclear what magnitude of this effect is in the context of network intrusion detection. The results of \cref{sec:results_classification_performance} indicate that label imbalance is, by far, the most important factor for the performance of DeepCASE. Regression over results derived from the datasets controlling for other characteristics that may affect decision boundaries shows that label imbalance outweighs other factors such as the size of the dataset or the heterogeneity of the sequences.

We envision two directions to increase robustness of alert post-processing methods such as DeepCASE in light of the imbalance inherent to the network intrusion detection domain: (1) reducing the imbalance in the data provided as input to such methods; and (2) making such methods more robust against imbalance. Tuning rulesets~\cite{vermeer2023-alert-alchemy} and increasing the quality of IDS rules are examples in the first direction. Researchers in the ML domain have already made steps towards the second direction by applying methods such as focal loss~\cite{lin2017}, cost-sensitive learning~\cite{lopez-13}, or by applying different methods for pre-processing alert data (e.g., constructing sequences).

Regardless of whether post-processing methods are used, deterministic alternatives can be utilized to reduce SOC workloads. Ruleset tuning has been discussed as a method to improve the performance of DeepCASE, but also reduces the number of irrelevant alerts by $99.6\%$, without the application of other post-processing methods. Hence, research on more traditional methods can still benefit ML methods for SOCs.

During our experiments, we computed various performance metrics, including the micro- and macro-averaged F1-scores in addition to the relaxed F1-score used in \cref{sec:results_classification_performance}. Some metrics like micro F1-score would not show significant differences even though the performance on the minority class was significantly affected. In light of the high imbalance, which is common in NIDS research, and the relative importance of the minority class, we consider micro-averaging inadequate for assessing NIDS or alert post-processing systems and encourage researchers to use appropriate performance measures~\cite{arp2022} that account for the imbalance and differences in cost associated with misclassifications between classes. We also encourage researchers to use robust metrics that account for rejection classes in a manner corresponding to the intended application context of the systems.

Previous research has stressed the importance of user studies in evaluating the explainability of decision support tools~\cite{nadeem2023}. The original work on DeepCASE suggested that vectors of total attention per event can serve as an explanation but has not evaluated its quality in that context. In \cref{sec:results_explainability} we make a first step towards evaluating the explainability of DeepCASE by involving two experts. Our results show that the classification for a non-negligible portion of vectors of total attention are poorly explained by DeepCASE. However, a reduction of the label imbalance through tuning resulted in a promising improvement in explanation correctness. Further research on the robustness of post-processing methods in the presence of imbalance can therefore enhance interpretability of these methods and encourage trust in SOC automation.

\noindent\textbf{Threats to validity.}
Data realism and representativeness are important issues~\cite{arp2022} and past research has raised concerns regarding the generation of benchmark datasets, specifically through artificial means~\cite{flood-24,engelen2021}.
Since the methods described in \cref{sec:experiment_design} and Appendix~\ref{app:control} introduce no new data, artificial diversity or incorrect labels have not been introduced in the tuned datasets.
We believe that injection of attack traffic as described in \cref{sec:data_provisioning} and incorporation of test data into the training set as described in \cref{sec:experiment_design} have not led to spurious correlations or data snooping threatening the validity of our conclusions. These pitfalls would artificially inflate the performance on the minority class, whereas the performance on the minority class remains distinctly lower than on the majority class, as shown in \cref{tab:cir} (Appendix~\ref{app:additional}). Moreover, these pitfalls are present equally in all performed experiments and do therefore not prevent comparisons between performance on different datasets to assess the effects of imbalance, although the mentioned performance need not be reflective of real-world deployments.
Note that datasets generated by means other than mere collection or tuning may not be representative of the real world but serve their purpose well in assessing the impact of effects in the data on the performance of alert post-processing methods. Therefore, the results of our control datasets should not be interpreted as reflecting real-world performance but only relative to the results on the datasets with different IR.
The choice of appropriate metrics in light of imbalanced data~\cite{arp2022} has been addressed in \cref{sec:metrics}.

We believe to have optimized DeepCASE well for the presence of label imbalance using the macro F1-score. Thereafter, we have used the relaxed F1-score, since it is considered more representative of the problem that a SOC would address using DeepCASE.
Although the macro F1-score used to optimize hyperparameters and the relaxed F1-score used to evaluate DeepCASE's performance are different, the macro and relaxed F1-scores are statistically correlated. Therefore, we deem it unlikely that optimizing for the macro F1-score has negatively affected conclusions drawn based on results using the relaxed F1-score.
When comparing the macro F1-scores shown in \cref{fig:5mf1} (Appendix~\ref{app:reproduce}) and the relaxed F1-scores shown in \cref{fig:5f1}, we observe that both metrics rank the performance of DeepCASE similarly for the different levels of imbalance for each of the tuned and control datasets used in the regression from \cref{sec:results_classification_performance}. We consider the relaxed F1-score to be more indicative of real-world performance and operational implications, since it accounts for the impact of alerts rejected (i.e., unclassified) by DeepCASE on time spent investigating by security analysts.

The regression presented in \cref{sec:results_classification_performance} shows a significant effect on the relaxed F1-score, despite the absence of other variables that we aimed to control for due to the high collinearity of those variables.
Since event IR and label IR are correlated, it is possible that (part of) the effect of label IR in the regression is actually the result of event IR instead. Despite this ambiguity, we remain confident that imbalance, be it event or label, has a significant effect on the classification performance.

Since expert 1 and expert 2 agreed with each other about as much as DeepCASE agreed with expert 1, we cannot draw strong conclusions from the correctness of DeepCASE's explanations. From feedback acquired from the experts on the provided explanations, the experts struggled the most with the ambiguous interpretation of attention assigned to events in the input sequence when no relevant events exist in the input sequence. Explanations offered by DeepCASE therefore remain a topic requiring future research, as it apparently is not trivial for experts to agree on what constitutes correct explanations.

\section{Conclusion} \label{sec:conclusion}
In this work, we evaluated the effect of label imbalance on DeepCASE and showed that classification performance increases when label imbalance is reduced as a result of tuning. Furthermore, we evaluated the correctness of explanations offered, and found that the correctness also improves by tuning. We found that tuning of rules in SOCs can reduce the imbalance and can effectively improve classification performance and correctness of explanations offered by alert post-processing methods like DeepCASE.

\section*{Data Availability}
\addcontentsline{toc}{section}{Data Availability}
To promote reproducibility, we archive an anonymized copy of the SOC dataset, the code used to create the tuned datasets and control datasets, and the code used to run and evaluate DeepCASE's performance at our institution.
The archived dataset and the code used in this research can be shared upon request with interested researchers.

\section*{Acknowledgment}
\addcontentsline{toc}{section}{Acknowledgment}
\looseness=-1Parts of this work were derived from a prior M.Sc. graduation project \cite{graduation-project}. The authors thank the unnamed SOC for their cooperation and for making data available for analysis, and Thijs van Ede discussing DeepCASE with us.
This publication is part of the CATRIN, INTERSECT, and SeReNity projects (with numbers {\small NWA.1215.18.003}, {\small NWA.1160.18.301}, and {\small CS.010}) which are (partly) financed by the Dutch Research Council (NWO). For the purpose of Open Access, a CC-BY 4.0 public copyright license is applied to any Author Accepted Manuscript version arising from this submission.

\bibliographystyle{IEEEtran}
\bibliography{IEEEabrv,main}

\appendices
\section{Hyperparameter optimization} \label{app:hyper}
DeepCASE has several hyperparameters that can be optimized as discussed in \cref{sec:deepcase} and detailed in \cref{tab:2hyper}. The first $1\%$ of the dataset that is selected only for hyperparameter optimization and is excluded from the remaining runs. Within this $1\%$ of the data, we make a $50/50$ split to obtain train and test sets for hyperparameter optimization. Similar to the approach described in \cref{sec:methodology}, we copy over \texttt{Incident} samples from the test set to the train set until both sets have similar proportions of \texttt{Incident} alerts.

The optimal parameters are selected using random grid search, where a point on the grid is randomly selected for trial, combining the advantages of grid search and manual search~\cite{bergstra2012} while avoiding an excessively large search space. Contrary to the original work, we optimize the macro F1-score instead of the micro F1-score~\cite{vanede2022} to favor performance on the minority class.

The hyperparameter optimization experiment is run before any data analysis is performed. Every time the dataset presented to DeepCASE changes, we first run a hyperparameter optimization experiment. This ensures DeepCASE can perform optimally for the dataset, such that the conclusions we reach from the data analysis are independent of the chosen hyperparameters, considering the optimal hyperparameters are typically different when a model is deployed on different data.

\begin{table*}[t]
    \centering
    \vspace{-0.10in}
    \caption{Overview of DeepCASE's hyperparameters.}
    \label{tab:2hyper}
    \vspace{-0.1in}
    \begin{tabular}{lp{0.7\linewidth}}
        \toprule
        Hyperparameter  & Function \\
        \midrule
        Context Length ($n$) & Maximum amount of events considered to be part of the context. \\
        Context Timeout ($t$) & Maximum duration between an event and the corresponding alert for the event to be considered part of the context. \\
        Context Builder Hidden Nodes & Amount of nodes that are part of the hidden layer of the context builder's neural network. \\
        $\delta$ (Label Smoothing Regularization) & Adjustment of target output label probabilities during training of context builder. \\
        $\tau_{confidence}$ (Confidence Threshold) & Minimum confidence required before an attention vector and the vector of total attention per event is used in the clustering algorithm. \\
        $\epsilon$ (DBSCAN Maximum Distance) & Maximum distance between vectors of total attention where the DBSCAN algorithm considers the different vectors to be part of the same cluster. \\
        Minimum Cluster Size & Minimum amount of vectors required to be considered a cluster, all within a maximum distance of $\epsilon$. \\
        \bottomrule
    \end{tabular}
    \vspace{-0.2in}
\end{table*}

\begin{table}[t]
    \centering
    \vspace{-0.05in}
    \caption{Possible values for DeepCASE's hyperparameters.}
    \label{tab:hyper_possible_values}
    \vspace{-0.1in}
    \begin{tabular}{lr}
        \toprule
        Hyperparameter  & Possible values \\
        \midrule
        Context Length ($n$) & $10$, $15$, $20$ \\
        Context Timeout ($t$)& $1$ day, $1$ week, $1$ month \\
        Context Builder Hidden Nodes & $32$, $64$, $128$ \\
        $\delta$ (Label Smoothing Regularization) & $[0, 1]$ with steps of $0.05$ \\
        $\tau_{confidence}$ (Confidence Threshold) & $[0.05, 1]$ with steps of $0.05$ \\
        $\epsilon$ (DBSCAN Maximum Distance) & $[0.05, 1]$ with steps of $0.05$ \\
        Minimum Cluster Size & $5$, $10$, $20$, $50$ \\
        \bottomrule
    \end{tabular}
    \vspace{-0.2in}
\end{table}

For discovering suitable hyperparameter values, we use random search on a grid with a limit on the amount of possible trials to find the optimal performing set of hyperparameters. The hyperparameters and their possible choices are shown in \autoref{tab:hyper_possible_values}. Random search is chosen because it generally outperforms grid search in both the time spent on computations and the performance of the model~\cite{bergstra2012}. However, if we rely purely on random search with the upper bound and lower bound of each parameter set to the highest and lowest values in the original paper, then the possible search space is so large ($409\,600\,000$ combinations) that an excessive number of trials would be required. Considering the required time and hardware, as well as the potential impact on the environment, we reduce the amount of possible combinations by randomly sampling from a grid, combining the advantages of random search with applying domain knowledge to ensure random search will avoid insensible parameters like a sequence length of 1. This approach significantly reduces the search space ($779\,760$ combinations), making hyperparameter optimization feasible.
The resulting optimal hyperparameters for each of the different sets are shown in \autoref{tab:hyper_values}.

\begin{table*}[t]
    \centering
    \vspace{-0.05in}
    \caption{Optimal Hyperparameters for the Different Datasets.}
    \label{tab:hyper_values}
    \vspace{-0.1in}
    \begin{tabular}{lrrrr}
        \toprule
        Hyperparameter      & Unfiltered    & High Label IR & Medium Label IR   & Low Label IR \\
        \midrule
        Context Length     & $15$            & $10$            & $20$                & $15$ \\
        Context Timeout    & $1$ Day         & $1$ Month       & $1$ Day             & $1$ Week \\
        Context Builder Hidden Nodes        & $64$            & $32$            & $32$                & $64$ \\
        $\delta$            & $0.25$          & $0.5$           & $0.35$              & $0$ \\
        $\tau_{confidence}$ & $0.05$          & $0.05$          & $0.05$              & $0.05$ \\
        $\epsilon$          & $0.8$           & $0.75$          & $0.9$               & $1$ \\
        Minimum Cluster Size& $50$            & $5$             & $5$                 & $5$ \\
        \bottomrule
    \end{tabular}
    \vspace{-0.2in}
\end{table*}

\section{Reproduction} \label{app:reproduce}
To further support the interpretation of our experiments and enable comparison with other datasets, we describe the relevant aspects of the data and make a comparison with the dataset used by Van Ede \textit{et al.}~\cite{vanede2022}. Respectively, we refer to these datasets as the SOC and Lastline datasets and their comparison is summarized in \cref{tab:datasets}.

\begin{table}[t]
    \centering
    \vspace{-0.05in}
    \caption{Properties of the SOC and Lastline datasets.}
    \label{tab:datasets}
    \vspace{-0.1in}
    \begin{tabular}{lrr}
        \toprule
         Property           &  SOC Dataset  & Lastline \\
         \midrule
         Fired Alerts       & $24.4$ million  & $10.5$ million\\
         Unique Rules       & $474$           & $291$\\
         Collection Time    & $5.5$ weeks     & $5$ months \\
         Distinct Hosts     & $1.3k$          & $388k$ \\
         Attacks            & $616$           & $45.1k$ \\
         \bottomrule
    \end{tabular}
    \vspace{-0.2in}
\end{table}

The SOC dataset contains twice the amount of fired alerts and nearly two times more distinct rules compared to the Lastline dataset, generated from less than 300 times as many hosts, and was gathered during a shorter time period. Based on these numbers, it seems that the environment from which the SOC dataset originates raises more FPs than that of the Lastline dataset, an observation explained by the fact that the rules used by the Suricata sensor generating the alerts had not been tuned by the cooperating SOC. This also implies that the SOC dataset has a higher imbalance and offers more complexity to post-processing methods such as DeepCASE.

We have opted to use only two labels for the SOC data:\textit{\texttt{Non-Incident}} and \textit{\texttt{Incident}}, where \texttt{Incident} alerts are alerts related to successful attacks and \texttt{Non-Incident} alerts are all alerts considered FPs. In the original work, DeepCASE was used to estimate threat levels associated with alerts, which can be considered a generalization of the task for which we employ it: distinguishing successful from unsuccessful attacks. Our less granular labeling presents DeepCASE with a less complex task.

One may raise concerns about the difference in the number of incidents between the Lastline dataset and the dataset used in this work.
We attribute this to a difference in the interpretation of what constitutes an incident and the corresponding labeling methodology. The cooperating SOC labels an alert as an attack if it is related to a malicious action that is \textit{successful}. Therefore, the SOC does not consider alerts related to malicious but unsuccessful actions as attacks. We suspect that in the Lastline dataset, the label \texttt{ATTACK} might also have been given to malicious but unsuccessful actions. As a result, the SOC dataset is more imbalanced. This high imbalance is in line with a previous characterization of SOC data~\cite{yang2024}, whose dataset was collected over four years from a real-world SOC and contained $115$ million alerts and only $227$ attacks, where their definition of ``attack'' aligns with ours.

As shown in \cref{fig:5mf1} (note the Figure also contains results on the control datasets described in Appendix~\ref{app:control}), the unfiltered dataset has a mean macro F1-score of $0.502$ for the IR experiment. We note that the Macro F1-scores obtained in the original work~\footnote{\url{https://github.com/Thijsvanede/DeepCASE/blob/sp/experiments/baseline/results/deepcase_after_automatic.txt\#L58}} are similar to the Macro F1-scores obtained in this work.
Thus, the performance of DeepCASE on the Lastline and SOC datasets is comparable despite differently defined attacks.

\begin{figure}[htb]
    \centering
    \vspace{-0.05in}
    \includegraphics[width=\linewidth]{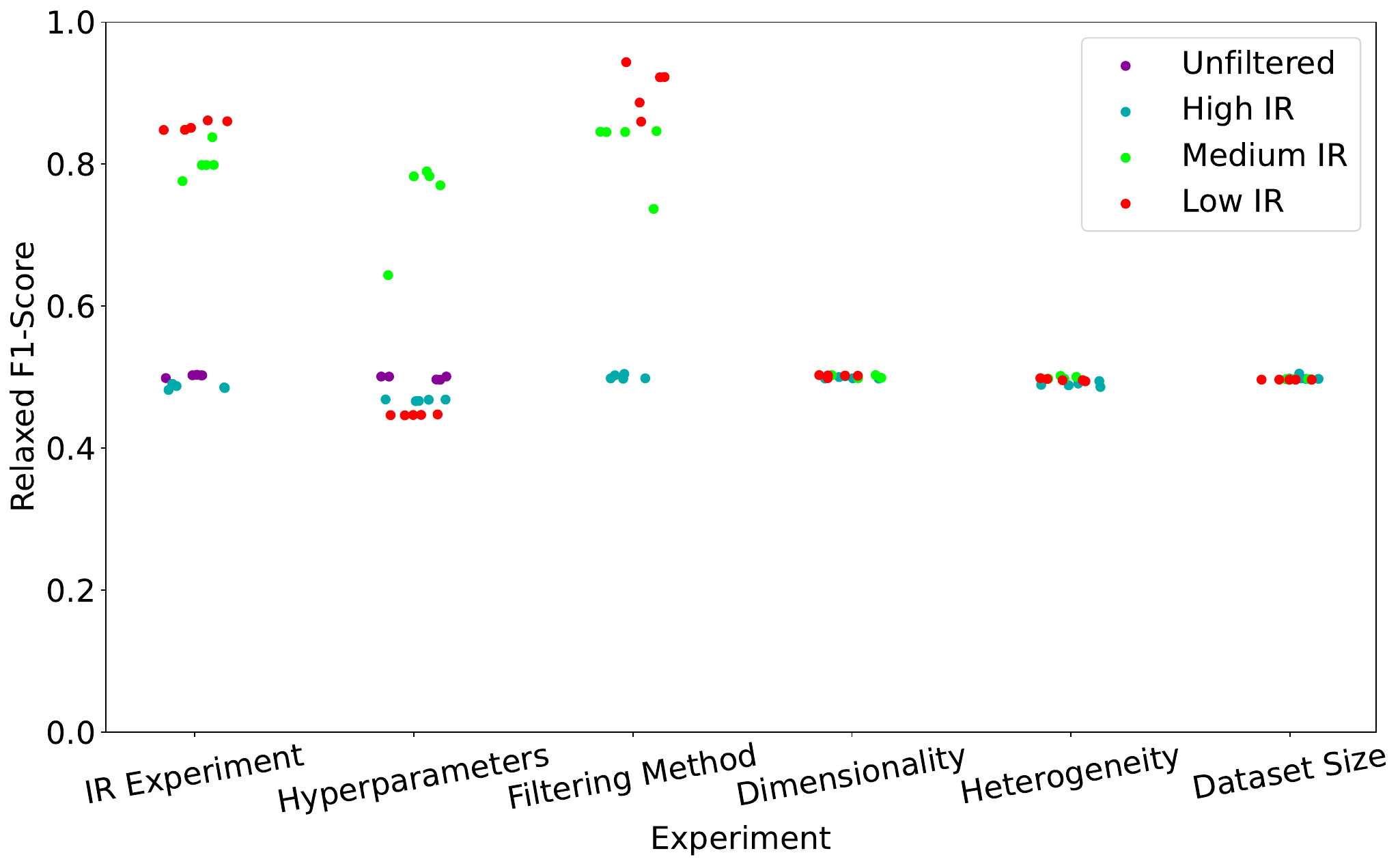}
    \vspace{-0.35in}
    \caption{Macro F1-score for All Experiments. Each point represents the Macro F1-score of one run.}
    \label{fig:5mf1}
    \vspace{-0.05in}
\end{figure}

Looking at the confusion matrix for the unfiltered dataset in \cref{tab:cir} of \cref{app:additional}, we can see that nearly all \texttt{Non-Incident} alerts are classified correctly, whereas nearly all \texttt{Incident} alerts are rejected. Among the few classified \texttt{Incident} alerts, most are classified incorrectly. Due to the imbalance in the dataset, the micro F1-score would almost only reflect the performance of the majority class and hence hide the poor performance on the minority class. This reaffirms our choice for the macro F1-score and relaxed F1-score discussed in \ref{sec:metrics}.
This poor performance on the \texttt{Incident} alerts is despite the copying of some \texttt{Incident} alerts from the test set to the train set as described in \cref{sec:methodology}.

The macro F1-score generally increases as label IR is reduced. The high IR dataset has a mean macro F1-score of $0.486$, the medium IR dataset has a mean macro F1-score of $0.802$ and the low IR dataset has a mean macro F1-score of $0.854$. This is in-line with our findings from \cref{sec:results_classification_performance} using the relaxed F1-score.

The macro F1-score for the low IR dataset for the hyperparameter control experiment described in Section~\ref{sec:control_experiment_hyperparameters} of Appendix~\ref{app:control} is surprisingly not as high as its IR counterpart, with a mean macro F1-score of $0.447$.
However, it seems to be the case that the default hyperparameters only work as expected for the unfiltered, high IR and medium IR datasets, due to their scores being similar with their IR experiment counterparts, with mean macro F1-scores of $0.499$, $0.468$ and $0.754$ respectively.
Since simply tuning the hyperparameters seems to increase the macro F1-score significantly for the low IR dataset, we can conclude that hyperparameter optimization for each dataset can significantly increase DeepCASE's performance.

\section{Control experiments and datasets} \label{app:control}
This appendix describes how we control for various experimental conditions such as hyperparameters (\cref{sec:control_experiment_hyperparameters}) and characteristics other than imbalance. Concretely, we describe how we derive control datasets in \cref{sec:control_datasets} and make some comparisons between datasets in \cref{sec:control_dataset_comparison}. The outcome of running DeepCASE controlling for these other characteristics is discussed in \cref{sec:results_classification_performance}.

\subsection{Hyperparameter control experiment}
\label{sec:control_experiment_hyperparameters}
While not a property of the dataset, we did identify different sets of hyperparameters (see Appendix~\ref{app:hyper}) for the different datasets, which are also different compared to the set of hyperparameters supplied by the original work~\cite{vanede2022}.
To verify whether the optimization is performed correctly and to assess the effect of hyperparameter tuning on DeepCASE, we perform a hyperparameter control experiment.
As such, we run the different datasets through DeepCASE using the selection of hyperparameters as suggested by the original work~\cite{vanede2022}.
The effect of hyperparameter optimization and the outcome of this control experiment is discussed in relation to the original work in Appendix~\ref{app:reproduce}.

\subsection{Control datasets}
\label{sec:control_datasets}
According to the methodology presented in \cref{sec:experiment_design_classification_performance}, we require control datasets to distinguish the effects of different characteristics that may affect classification performance.
The characteristics we control for are: dataset size, event imbalance, context heterogeneity, and context builder input dimensionality.
In the following, we describe the derivation of four types of control datasets, each intended to vary the aforementioned characteristics w.r.t. the unfiltered dataset as depicted in Table~\ref{tab:control_datasets_characteristics}. We generate a control dataset of each type for every tuned dataset to match some of its properties and vary other characteristics. Runs on the control dataset use the same hyperparameter values as used for the corresponding tuned dataset.

\begin{table*}[t]
    \centering
    \caption{Comparison of expected effects of tuned datasets and control datasets on controlled characteristics with respect to the unfiltered dataset.}
    \label{tab:control_datasets_characteristics}
    \vspace{-0.1in}
    \begin{tabular}{lrrrrr}
        \toprule
         \textbf{Control Dataset} & \textbf{Label Imbalance} & \textbf{Event Imbalance} & \textbf{Dataset Size} & \textbf{Heterogeneity} & \textbf{Dimensionality} \\
         \midrule
        \textbf{SOC Ruleset Tuning} & - & - & - & + & - \\
        \textbf{Filtering Method} & - &  & - &  & - \\
        \textbf{Dataset Size} &  &  & - &  & - \\
        \textbf{Dimensionality} &  &  &  &  & - \\
        \textbf{Heterogeneity} &  & - &  & + & \\
        \bottomrule
    \end{tabular}
    \vspace{-0.20in}
\end{table*}

\subsubsection{Filtering Method}
To derive this control dataset, we filter out sequences until we match the dataset size of the corresponding tuned dataset. Sequences are selected at random with uniform probability of selecting a specific sequence, with the exception that sequences with the \texttt{Incident} label are never removed.
As a result, the label imbalance and dataset size are reduced from the unfiltered dataset. By chance, all sequences associated with a rule are randomly selected to be filtered for many infrequent rules, reducing the dimensionality as a result.
Contrary to the tuned datasets, the event IR and heterogeneity remain largely unaffected.

\subsubsection{Dataset Size}
To derive this control dataset, we follow the same approach as for the Filtering Method experiment but also aim to fix the label imbalance to be equal to that of the unfiltered dataset. This allows us to better understand the effects of dataset size on classification performance. To this end, we perform an additional step where we filter out sequences with the \texttt{Incident} label uniformly at random to increase the label IR until it matches that of the unfiltered dataset.
As a result, the dataset size is reduced from the unfiltered dataset. Similarly to the Filtering Method control dataset, the dimensionality is also reduced as a result.
Contrary to the tuned datasets, label IR, event IR, and heterogeneity of these control datasets remain similar to those of the unfiltered dataset.

\subsubsection{Dimensionality}
With context builder input dimensionality, we refer to the amount of unique alerts in the dataset. Filtering out a rule will remove all alerts generated by the said rule from the dataset, which affects multiple contexts in the data and reduces how many unique contexts can be constructed.
To derive this control dataset, we randomly filter out alerts with probabilities inversely related to the frequency of alerts in the data. The probability of filtering out an alert $i$ is shown in \cref{eq:rq1-dimensionality}, where $a_c$ is the number of alerts in the dataset and $a_i$ is the number of occurrences of the alert $i$. The probability is normalized to sum to 1. The square is included to encourage more aggressive, and hence faster, filtering. These probabilities make it more likely that infrequent alerts are filtered out and less likely that common alerts are filtered. Note that sequences with the \texttt{Incident} label are never removed.
The dimensionality of the resulting control dataset is reduced from the unfiltered dataset to match that of the corresponding tuned dataset.
Contrary to the tuned datasets, label IR, event IR, dataset size, and heterogeneity of these control datasets remain similar to those of the unfiltered dataset.

\begin{equation}
\left( \frac{a_c}{a_i} \right) ^2
\label{eq:rq1-dimensionality}    
\vspace{-0.05in}
\end{equation}

\subsubsection{Heterogeneity}
Context heterogeneity refers to how heterogeneous the contexts of a dataset are. The more unique contexts a dataset contains, the more heterogeneous the dataset is.
To assess the effect of heterogeneity, we introduce another control dataset with the aim of only varying the heterogeneity and not other characteristics such as label imbalance, dimensionality, or dataset size.
As such, we randomly filter out common sequences until we match the desired context heterogeneity of the corresponding tuned dataset. The probability of filtering out a sequence $i$ is shown in \cref{eq:rq1-hetero}, where $s_c$ is the number of contexts in the dataset and $s_i$ is the amount of occurrences of context $i$. The probability is normalized to sum to 1. The square is included to encourage more aggressive, and hence faster, filtering. We perform upsampling to match the dataset size of the unfiltered dataset.
The resulting control dataset has the same heterogeneity as its tuned counterpart (hence increased from the unfiltered dataset) but label imbalance and dimensionality similar to that of the unfiltered dataset. The event IR is also reduced from the unfiltered dataset as a result of the filtering.

\begin{equation}
\left( \frac{s_i}{s_c} \right) ^2
\label{eq:rq1-hetero}    
\vspace{-0.05in}
\end{equation}

\subsection{Filtering Method Control Dataset Comparison with Tuned Datasets}
\label{sec:filtering_method_comparison}
\label{sec:control_dataset_comparison}
Looking deeper into the Filtering Method (FM) control experiment specifically and its comparison to the tuned datasets which is summarized in \cref{tab:control_datasets_characteristics}, we see that all controlled characteristics are significantly different compared to the tuned datasets, except for the label IR. For all FM control experiment datasets, the dimensionality, and heterogeneity are lower than in the corresponding tuned datasets. These properties seem to indicate that the composition of the filtering method control dataset is more homogeneous and less varied w.r.t. the event types than the tuned datasets. This is also reflected in the number of unique contexts, which are generally lower for the filtering method experiment compared to the IR experiment. This corresponds to the intuition that if you give a neural network largely similar data, its ability to classify those data correctly improves, hence explaining the significant difference in performance.

\section{Expert labeling validation} \label{app:expert}
Another expert (also recruited from the SOC, and referred to as expert 2) independently assessed a random selection of the same vectors of total attention per event, the size of $11\%$ of the total set in-line with the approach described in \cref{sec:experiment_design_explainability}.

\begin{figure}[b]
    \centering
    \vspace{-0.15in}
    \includegraphics[width=\linewidth]{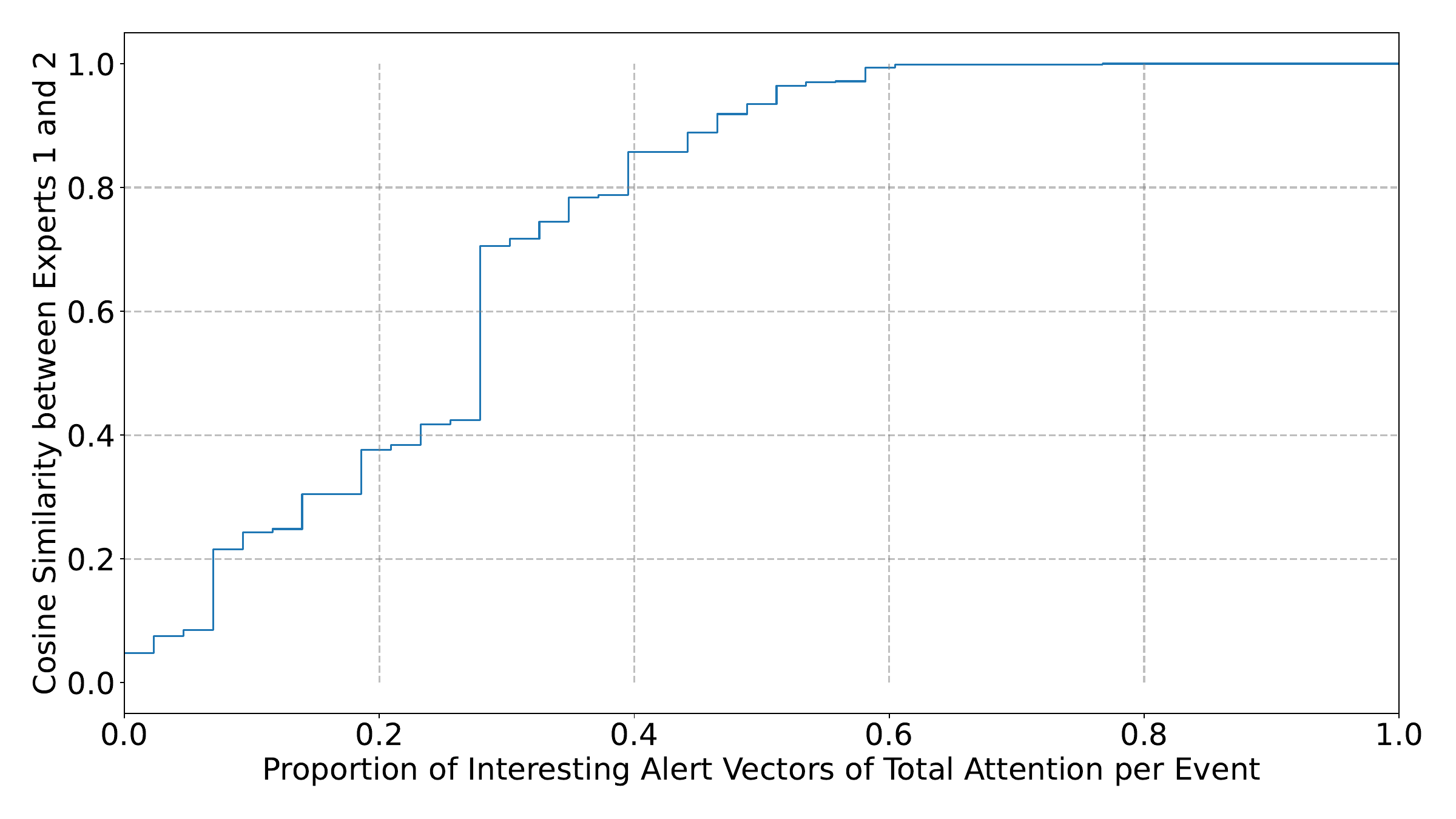}
    \vspace{-0.35in}
    \caption{Cumulative Distribution of the Cosine Similarity of Expert Created vectors of total attention per event between Expert 1 and 2}
    \label{fig:6expert}
    \vspace{-0.05in}
\end{figure}

When comparing the manually created vectors of total attention per event, as shown in the form of a cumulative distribution in \cref{fig:6expert}, we see that the experts have created similar vectors for the majority of the set. More specifically, for approximately $40\%$ of the compared vectors, the labeling of both experts are identical. More than $60\%$ of vectors have a cosine similarity higher than $0.8$, suggesting high agreement between experts on what constitutes correct labeling. Only about $20\%$ of vectors have a cosine similarity below $0.4$, suggesting a low agreement. Considering that expert 2 has independently created similar vectors of total attention per event for the majority of the randomly selected vectors, we conclude that the ground truth created by expert 1 provides a good baseline for which events in the input sequence should be assigned attention.

\section{Additional results} \label{app:additional}
This appendix lists confusion matrices from our experiments described in \cref{sec:methodology} and \cref{sec:results}. \cref{tab:cir} reports on the performance on the unfiltered and tuned datasets. \cref{tab:cfm} reports on the performance on the filtering method control datasets.

As explained in \cref{sec:results_classification_performance}, the difference in performance between runs of DeepCASE on the tuned datasets and the Filtering Method (FM) control datasets is attributed to the lower data complexity of the FM control datasets.
When comparing the aggregated classification reports on the tuned datasets and FM control datasets, shown in \cref{tab:cir}  and \cref{tab:cfm}, we see this reflected in the results.
When comparing the performance on the tuned datasets with the performance on the FM control datasets, both the mean amount and the standard deviation of \texttt{Non-Incident} alerts that are not classified are consistently lower for the FM control datasets, with a mean of $5.0\%$ versus $0.8\%$ for the high IR dataset, $11.0\%$ versus $1.1\%$ for the medium IR dataset and $14.9\%$ versus $1.3\%$ for the low IR dataset.
Moreover, both the mean amount and the standard deviation of misclassified \texttt{Incident} alerts is also about the same, indicating that the number of alerts investigated on the FM control datasets is smaller than on the tuned datasets.
Since the relaxed F1-score punishes rejected alerts, the lower amount of rejected \texttt{Non-Incident} alerts explains why the relaxed F1-score is higher on the FM control datasets.
The runs of DeepCASE on these control datasets show that DeepCASE performs better on datasets that are largely repetitive in nature.

\begin{table*}[htb]
    \centering
    \vspace{-0.10in}
    \caption{Aggregated Confusion Matrix for the unfiltered and tuned datasets.}
    \label{tab:cir}
    \vspace{-0.1in}
{
    \begin{tabular}{llrrrrrr}
        \toprule
       Dataset & \multicolumn{1}{r}{Predicted} & \multicolumn{2}{c}{Not Classified} & \multicolumn{2}{c}{\texttt{Non-Incident}} & \multicolumn{2}{c}{\texttt{Incident}} \\
       & True & $\mu$ & $\sigma$ & $\mu$ & $\sigma$ & $\mu$ & $\sigma$ \\
       \midrule
       \multirow{4}*{Unfiltered} & \texttt{Non-Incident}  & $104133$ & $5369$ & $19294023$ & $5235$ & $183$ & $159$  \\
& & $0.5\%$ & & $99.5\%$ & & $0.0\%$ & \\
& \texttt{Incident} & $610$ & $1$ & $3$ & $1$ & $2$ & $1$  \\
& & $99.2\%$ & & $0.5\%$ & & $0.3\%$ & \\[1.5ex]
       \multirow{4}*{High IR} & \texttt{Non-Incident}  & $171200$ & $19912$ & $3202255$ & $30326$ & $21643$ & $24987$  \\
& & $5.1\%$ & & $94.3\%$ & & $0.6\%$ & \\
& \texttt{Incident} & $606$ & $0$ & $7$ & $2$ & $3$ & $2$  \\
& & $98.4\%$ & & $1.1\%$ & & $0.5\%$ & \\[1.5ex]
       \multirow{4}*{Medium IR} & \texttt{Non-Incident}  & $23855$ & $699$ & $193760$ & $699$ & $57$ & $0$  \\
& & $11.0\%$ & & $89.0\%$ & & $0.0\%$ & \\
& \texttt{Incident} & $281$ & $30$ & $0$ & $0$ & $334$ & $30$  \\
& & $45.7\%$ & & $0.0\%$ & & $54.3\%$ & \\[1.5ex]
       \multirow{4}*{Low IR} & \texttt{Non-Incident}  & $11642$ & $275$ & $66395$ & $275$ & $61$ & $2$  \\
& & $14.9\%$ & & $85.0\%$ & & $0.1\%$ & \\
& \texttt{Incident} & $174$ & $11$ & $0$ & $0$ & $441$ & $11$  \\
& & $28.3\%$ & & $0.0\%$ & & $71.7\%$ & \\
        \bottomrule
    \end{tabular}
}
    \vspace{-0.15in}
\end{table*}

\begin{table*}[htb]
    \centering
    \vspace{-0.10in}
    \caption{Aggregated Confusion Matrix for the Filtering Method (FM) control datasets.}
    \label{tab:cfm}
    \vspace{-0.1in}
{
    \begin{tabular}{llrrrrrr}
        \toprule
       Dataset & \multicolumn{1}{r}{Predicted} & \multicolumn{2}{c}{Not Classified} & \multicolumn{2}{c}{\texttt{Non-Incident}} & \multicolumn{2}{c}{\texttt{Incident}} \\
       & True & $\mu$ & $\sigma$ & $\mu$ & $\sigma$ & $\mu$ & $\sigma$ \\
       \midrule
       \multirow{4}*{High FM} & \texttt{Non-Incident}  & $27665$ & $3407$ & $3366150$ & $3086$ & $1407$ & $2681$  \\
       & & $0.8\%$ & & $99.1\%$ & & $0.1\%$ & \\
       & \texttt{Incident} & $ 608$ & $ 2$ & $ 4$ & $ 1$ & $ 2$ & $ 2$  \\
       & & $99.0\%$ & & $0.7\%$ & & $0.3\%$ & \\[1.5ex]
       \multirow{4}*{Medium FM} & \texttt{Non-Incident}  & $ 2321$ & $ 174$ & $ 215473$ & $ 174$ & $ 1$ & $ 0$  \\
       & & $1.1\%$ & & $98.9\%$ & & $0.0\%$ & \\
       & \texttt{Incident} & $ 311$ & $ 52$ & $ 1$ & $ 1$ & $ 302$ & $ 53$  \\
       & & $50.7\%$ & & $0.1\%$ & & $49.2\%$ & \\[1.5ex]
       \multirow{4}*{Low FM} & \texttt{Non-Incident} & $ 1035$ & $129$ & $ 77160$ & $182$ & $28$ & $56$  \\
       & & $1.3\%$ & & $98.6\%$ & & $0.1\%$ & \\
       & \texttt{Incident} & $ 167$ & $33$ & $0$ & $0$ & $447$ & $33$  \\
       & & $27.2\%$ & & $0.0\%$ & & $72.8\%$ & \\
        \bottomrule
    \end{tabular}
}
    \vspace{-0.15in}
\end{table*}

\end{document}